\documentclass{PoS}

\def\Dsl{\hbox{/\kern-.6000em D}} 

\def\dsl{\,\raise.15ex\hbox{/}\mkern-13.5mu D}

\def\psip#1{\psi_{\mathbf{#1}}}
\def\chip#1{\chi_{\mathbf{#1}}}

\def\ltap{\ \raise.3ex\hbox{$<$\kern-.75em\lower1ex\hbox{$\sim$}}\ }
\def\gtap{\ \raise.3ex\hbox{$>$\kern-.75em\lower1ex\hbox{$\sim$}}\ }
\def\OMIT#1{}

\def\lsim{\mathrel{\raise.3ex\hbox{$<$\kern-.75em\lower1ex\hbox{$\sim$}}}}
\def\gsim{\mathrel{\raise.3ex\hbox{$>$\kern-.75em\lower1ex\hbox{$\sim$}}}}

\newcommand{\bmk}{\mathbf k}
\newcommand{\bmp}{\mathbf p}

\newcommand{\bmr}{\mathbf r}

\newcommand{\bmD}{\mathbf D}
\newcommand{\bmS}{\mathbf S}

\vbox{ \hbox{MPP-2006-40} 
}

\title{Top Threshold Physics}

\ShortTitle{Top Threshold Physics}

\author{\speaker{Andr\'e H. Hoang}
\\
        MPI Munich (Germany)\\
        E-mail: \email{ahoang@mppmu.mpg.de}}

\abstract{
Running a future Linear Collider at the top pair threshold allows
for precise measurements of the mass, the widths and the couplings of
the top quark. I give a nontechnical review on recent theoretical
developments and the theory status in top threshold physics concerning
QCD corrections and top quark finite lifetime and
electroweak effects. I also discuss threshold physics in the context of
measurements of the top Yukawa coupling from $e^+e^-\to t\bar t H$ and of
squark pair production. 
}

\FullConference{International Workshop on Top Quark Physics\\
		 January 12-15, 2006\\
		 Coimbra, Portugal}

\begin{document}

\section{Introduction}
\label{sectionintroduction}

The measurement of the total $t\bar t$ cross section constitutes a
major part of the top physics program at a future $e^+e^-$ Linear
Collider (LC). From the location of the rise of the cross section
measurements of the top quark mass will be gained, while from the shape
and the normalization one can extract the top quark width and get information
on the top Yukawa coupling and the strong coupling. I will begin this talk
with some comments on these measurement to provide an understanding of the
requirements that are imposed on the theoretical predictions. 

\begin{figure*}[htb]
\begin{center}
{\includegraphics[bb=0 10 567 350,width=7.5cm]{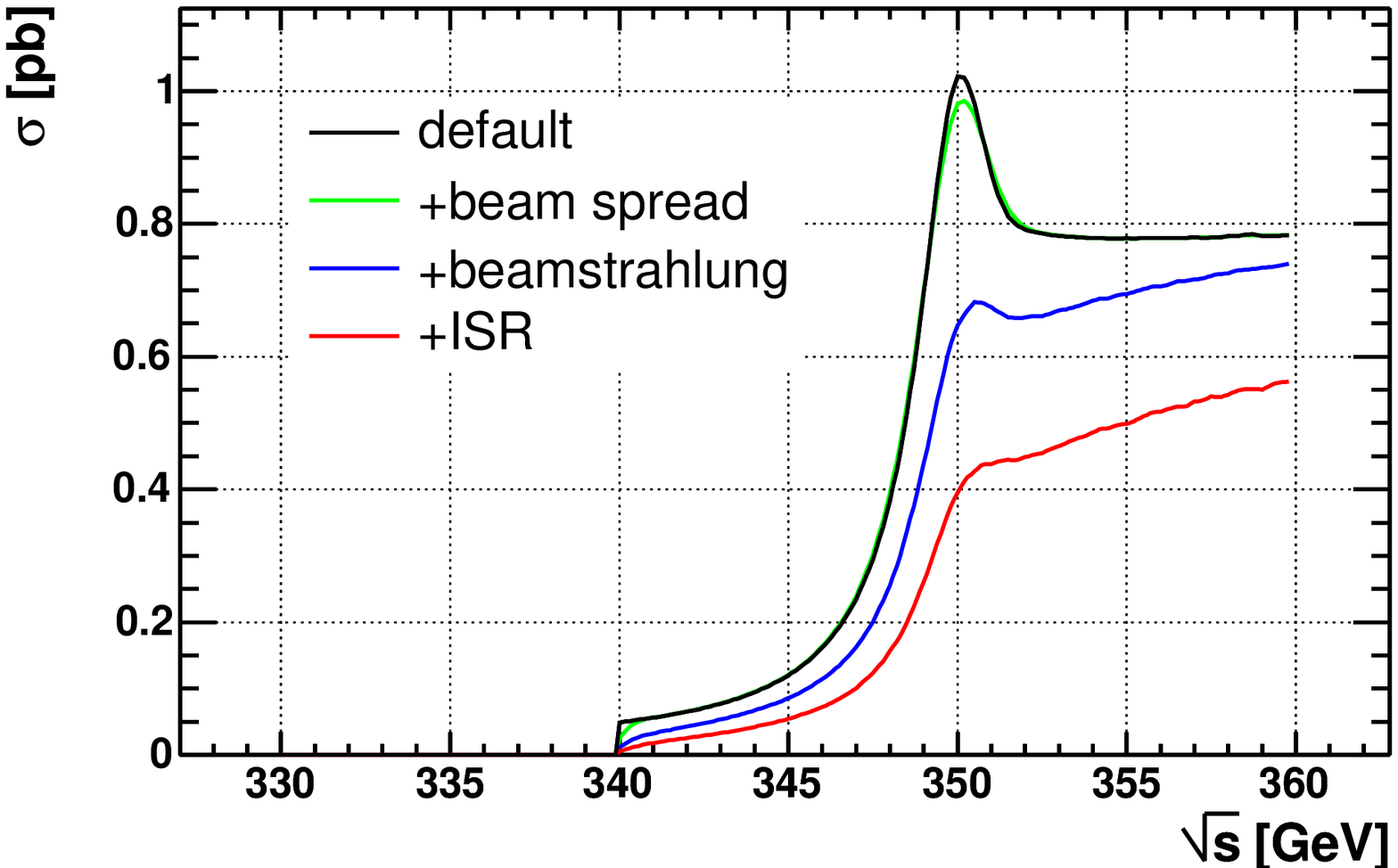}
\includegraphics[bb=0 10 567 350,width=7.5cm]{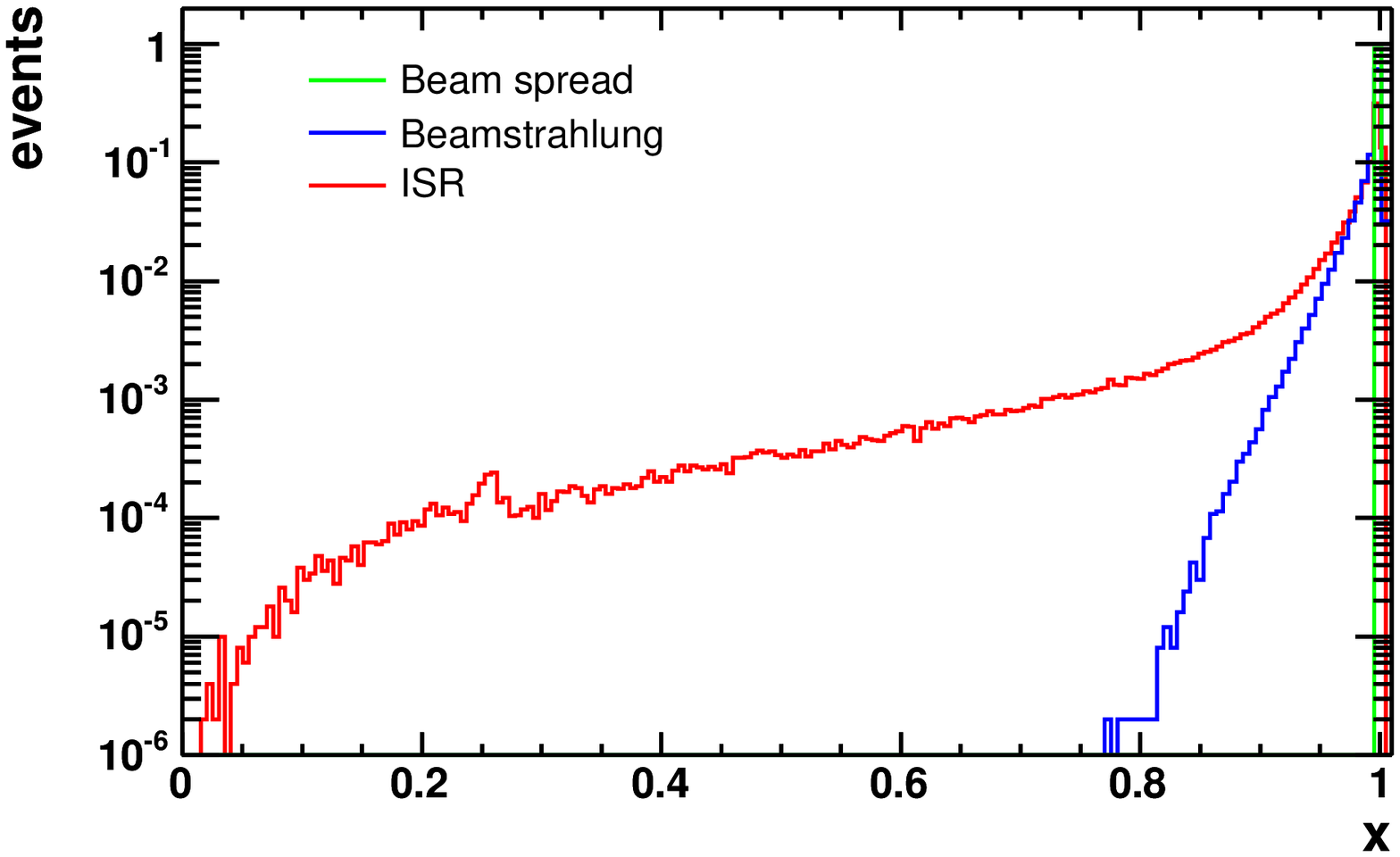}}
\caption{Left: Smearing of the "partonic" $t\bar t$ cross section
  by beam effects and initial state radiation.  Right
  panel: Simulation of beam spread, beamstrahlung and ISR as
  distributions of $x=\sqrt{s}/\sqrt{s_0}$ (where $\sqrt{s_0}$ is the nominal
  c.m. energy of the machine). The figures are from
  Refs.~\cite{Boogert:2002jr,boogerttalk}} 
\label{figlumi}
\end{center}
\end{figure*}

With a luminosity of a few times $10^{34}/(cm^2 s)$ experimental
simulation assume a total luminosity of several $100~fb^{-1}$ being spent
for a full threshold scan. Distributed into 10 scan points this corresponds 
to about $10^4$ $t\bar t$ pairs for those c.m.\,energies where the cross
section reaches the $1~pb$ level. To determine the observed
experimental cross section $\sigma_{t\bar t}^{\rm obs}$ one needs to
have a very good knowledge on the luminosity spectrum ${\cal L}(x)$
which accounts for the machine-dependent beam energy spread, the
effects of beamstrahlung and initial state
radiation~\cite{Boogert:2002jr,Cinabro1}, 
\begin{eqnarray}
\sigma_{t\bar t}^{\rm obs}(\sqrt{s}) &  = &
\int_0^1\,{\cal L}(x)\,\sigma^0_{t\bar t}(x^2\sqrt{s})
\,,
\label{convolution}
\end{eqnarray}
where $\sigma^0_{t\bar t}$ is the "partonic" cross section without
initial state beam effects. The luminosity spectrum leads to a smearing
of the partonic cross section and to a reduction of the observed cross
section (see Fig.~\ref{figlumi}). 

The c.m.~\,energy where the cross section rises is strongly related to
twice the top quark mass. The experimental statistical uncertainty in
top quark mass measurements is around $20$~MeV~\cite{TTbarsim} and
there is also an effect from uncertainties in the knowledge of ${\cal
  L}(x)$ which affects the top quark mass by probably less than
$50$~MeV~\cite{boogerttalk}. In contrast to the mass reconstruction
method that 
is traditionally used at the hadron colliders there is a very good
knowledge on the intrinsic theoretical uncertainties in this type of top quark
mass measurement. This is because the cross section line-shape can be
computed precisely with perturbative methods. One can rely on perturbative
methods because the rather large top quark width $\Gamma_t\approx
1.5$~GeV suppresses non-perturbative effects and prevents the
formation of toponium bound states. So the lineshape can be computed
as a function of the Lagrangian top quark mass in any given scheme
without ambiguities. The facts that we are considering a $t\bar t$
color singlet state and that one, in principle, just needs to count $t\bar t$
event (in the experimental measurement)
simplifies the task a lot. From NNLL order QCD computations
and from general arguments based on studies of QCD perturbation theory
at high orders~\cite{renormalons} it is known that the best
perturbative 
stability in the c.m.~\,energy where $\sigma^0_{t\bar t}$ rises is
obtained in so-called top threshold mass schemes~\cite{synopsis} such
as the 1S mass~\cite{1Smass,HT1S}, which I will use for the rest of this
presentation. This 
means that the top quark mass that is measured from the scan is a top
threshold mass, such as the 1S mass. An important issue here is that
threshold masses can be related reliably to e.g. the
$\overline{\mbox{MS}}$ mass (see e.g. Ref.~\cite{HT1S}) that is
frequently used for new physics studies or electroweak precision
observables. The top quark pole mass is known to lead to a much worse
higher order behavior~\cite{synopsis}, but it never becomes relevant
in these considerations anyway, except as maybe a useful quantity at
intermediate steps of the computations. Simulations have shown that the 1S
mass can be 
determined with theoretical uncertainties of about
$100$~MeV~\cite{Peralta1}. 

The top quark couplings and its total width can be determined from the
normalization of the cross section and the details of the line-shape
form. The strong coupling and the Yukawa coupling affect the
attraction of the $t\bar t$ pair and determine the normalization. The
top width determines the sharpness of the peak in $\sigma^0_{t\bar t}$. For 
$300~fb^{-1}$ distributed over 10 scan points the experimental errors
are smaller than $50$~MeV for the top width and at the level $0.001$
for $\alpha_s(M_Z)$~\cite{TTbarsim}. If the mass of the standard Higgs
is close to the 
present lower experimental bound, the Yukawa coupling to the top can be 
measured with around $35\%$ precision. To achieve comparable
theoretical errors the normalization and the line-shape form need to
be known with a precision of better than $3\%$. As we will see below
there is still some work to be invested to reach this goal.

The physics at the top threshold involves a number of nontrivial theoretical
issues related to the non-relativistic dynamics and  the finite top quark
lifetime that need to be addressed all at the same time. Gluon exchange
leads to singular terms $\propto (\alpha_s/v)^n$ and $\propto(\alpha_s\ln v)^n$
in $n$-loop perturbation theory where $v\ll 1$ is the top
velocity. The singularities enforce the parametric (power) counting
$v\sim\alpha_s\ll 1$, i.e.\,one needs to expand simultaneously in
$\alpha_s$ and $v$, and the use of an effective field theory (EFT) to
sum the singular terms up to all orders in $\alpha_s$. Due to the large top quark
lifetime these computations can be carried out perturbatively based on
the counting just mentioned. Interestingly the top width is
approximately equal to the typical top kinetic energy $\Gamma_t\sim
m_t\alpha\sim E_{\rm kin}\sim m_t\alpha_s^2$, so the effects of the
top lifetime cannot be treated as a perturbation and need to be
implemented systematically starting from the LL approximation. Due to the
relation between width and kinetic energy the combined expansion is
based on the parametric counting
\begin{equation}
v\sim\alpha_s\sim \alpha^{1/2} \ll 1
\,.
\label{powercounting}
\end{equation}
At this point a number of additional subtleties arise since the very notion of
the theoretical $t\bar t$ cross section requires a careful definition that
accounts e.g.\,also for the experimental information how the cross section is
being measured. 
In the following sections I discuss the status of the theoretical
predictions for the total cross section $\sigma^0_{t\bar t}$ concerning QCD 
(Sec.~\ref{sectionqcd}) and finite lifetime and electroweak effects
(Sec.~\ref{sectionew}). In
Secs.~\ref{sectionyukawa} and \ref{sectionsquark} I discuss
applications of these theoretical tools to Yukawa coupling
measurements from $e^+e^-\to t\bar tH$ and to squark pair production.

\section{QCD Effects}
\label{sectionqcd}

\begin{figure*}[htb] %
\begin{center}
\includegraphics[bb=230 570 428 710,width=4cm]{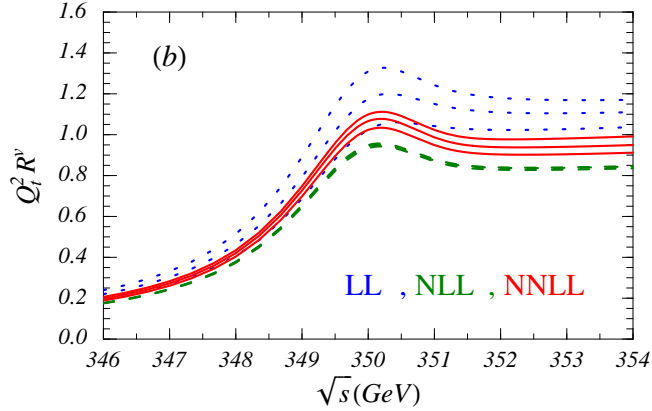}
\vskip  2.3cm
 \caption{
 Predictions for $R$ in renormalization group improved perturbation theory at
 LL (dotted lines), NLL (dashed lines) and NNLL (solid lines) order. 
 For each order curves are plotted for $\nu=0.15$, $0.20$, and $0.3$. 
 The effects of the luminosity spectrum are not included.~\cite{HoangEpiphany}}
\label{figttnnll} 
 \end{center}
\end{figure*}

Schematically the perturbative expansion and summations for the 
cross section have the form 
\begin{equation} R \, =\, \frac{\sigma_{t\bar t}}{\sigma_{\mu^+\mu^-}} 
\, = \,
 v\,\sum\limits_{k,i} \left(\frac{\alpha_s}{v}\right)^k
 \left(\alpha_s\ln v \right)^i \,\times \,\left\{
 1\,\,\mbox{(LL)} \,,\,\, 
 \alpha_s, v\,\, \mbox{(NLL)}\,,\,\, 
 \alpha_s^2, \alpha_s v, v^2\,\,\mbox{(NNLL)}
\right\}
\,.
\label{scheme}
\end{equation}
where the terms $\alpha_s/v$ and $\alpha_s\ln v$ are counted of order
$1$. The singular terms originate from ratios of the physical scales
$m_t$ (hard), $\bmp_t\sim m_t v$ (soft) and $E_t\sim m_t v^2$
(ultrasoft). The summations are achieved systematically in the
various orders of approximation by construction of a low energy EFT,
generically called nonrelativistic QCD (NRQCD), that describes correctly the 
nonrelativistic fluctuations of full QCD for the kinematic situation
of the top quarks close to threshold. A number of different versions
of NRQCD exist~\cite{BBL,pNRQCD,LMR,NRQCDc}, each of which aiming (in
principle) on 
applications in different physical situations. The EFT vNRQCD
(``velocity''NRQCD)~\cite{LMR,ultrasoft} has been designed for
predictions  at the $t\bar t$  threshold in the
scheme~(\ref{scheme}). It treats the case  
$m_t\gg \bmp_t\gg E_t > \Lambda_{\rm QCD}$, i.e. all physical scales
are perturbative, but also has the correlation $E_t=\bmp_t^2/m_t$ built
in at the field theoretical level.\footnote{
This differs from the pNRQCD approach~\cite{pNRQCD} where $\bmp_t$ and $E_t$
and the corresponding renormalization and matching scales are independent at
the field theoretical level. The correlation required for the description of
the $t\bar t$ dynamics~\cite{ManoharSoto1} is implemented by hand, see
also~\cite{Signertalk}.  
With some care applied, the predictions of both approaches should agree.
} The latter is important to achieve
the correct summation of the logarithmic $\ln v$ terms by renormalization
group evolution. The EFT consists of a Lagrangian with local operators
made from top and gluon fields that describe the quantum fluctuations that  
are resonant at the nonrelativistic scales $\bmp_t$ and $E_t$. High energy
fluctuations that occur in EFT loop diagrams and off-shell fluctuations are
accounted for in the EFT renormalization procedure and by matching the EFT
coefficients to the full theory at the hard scale $m_t$. This fixes the
matching (initial) conditions and the renormalization group running of the
coefficients. The large logarithms $\ln v$ are summed by evolving the
coefficient to the low-energy scale such that all large logs disappear from
the EFT matrix elements. At LL order the EFT Lagrangian relevant for
$e^+e^-\to t\bar t$ at threshold has the simple form 
\begin{eqnarray}
 \label{Lke}
 {\mathcal L}(x) &=& \sum_{\bmp}
   \psip{\bmp}^\dagger   \biggl\{ i D^0 - {(\bmp-i\bmD)^2 \over 2 m_t} 
   +\frac{{\bmp}^4}{8m_t^3}  
   + \frac{i}{2} \Gamma_t \bigg( 1 - \frac{{\bmp}^2}{2 m_t^2} \bigg) 
   - \delta m_t(\nu) \biggr\} \psip{\bmp} + (\psip{\bmp} \to\chip{\bmp})
\nonumber\\ &&
-\,\sum_{{\bmp},{\bmp^\prime}}
 V({\bmp,\bmp^\prime})\,\psi_{\bmp^\prime}^\dagger \psi_{\bmp}
   \chi_{-\bmp^\prime}^\dagger \chi_{-\bmp}
\,,
\nonumber\\[1mm]
 V({\bmp},{\bmp^\prime}) & = & 
 \frac{\mathcal{ V}_c(\nu)}{\bmk^2}
 + \frac{\mathcal{ V}_k(\nu)\pi^2}{m|{\bmk}|}
 + \frac{\mathcal{ V}_r(\nu)({\bmp^2 + \bmp^{\prime 2}})}{2 m_t^2 \bmk^2}
 + \frac{\mathcal{ V}_2(\nu)}{m_t^2}
 + \frac{\mathcal{ V}_s(\nu)}{m_t^2}{\bmS^2}\,,\quad (\bmk=\bmp^\prime-\bmp)
\label{vNRQCDpotential}
\end{eqnarray}
where $\psip{\bmp}$ and $\chip{\bmp}$ destroy top and
antitop quarks with momentum ${\bmp}$ and
$D^\mu$ is the covariant derivative with respect to ultrasoft gluons;
the term $V({\bmp,\bmp^\prime})$ contains the Coulomb potential. I have also
shown a few terms that come in at NNLL. All couplings are
functions of the dimension-zero renormalization group scaling parameter
$\nu$. (Please note the subtle difference between the scale 
$\nu$ ({\tt \$$\backslash$nu\$}) and the velocity $v$ ({\tt \$v\$}) for the fonts
used in these proceedings.) 
At $\nu=1$ (hard scale $m_t$) the coefficients are determined from the
matching procedure and at $\nu\sim v\sim\alpha_s$ the EFT matrix elements are
computed. The scaling from $\nu=1$ to $\alpha_s$ sums all logs of ratios of the
scales $m_t$, $\bmp_t$ and $E_t$ and also accounts for the correlation of
$\bmp_t$ and $E_t$~\cite{LMR}. The coefficient $\delta m_t$ is determined (unambiguously
order by order) by the mass definition that is used. The term 
$\psip{\bmp}^\dagger i\Gamma_t \psip{\bmp}$
accounts for the top  decay at LL order, which I will discuss in more detail
in the next section.  
The equation of motion for the $t\bar t$ system obtained from Eq.~(\ref{Lke})
is a Schr\"odinger equation. At NNLL order QCD it has, in configuration space, 
the form 
\begin{equation}
\bigg(\,
-\frac{{\mbox{\boldmath $\nabla$}}^2}{m_t^2} 
- \frac{{\mbox{\boldmath $\nabla$}}^4}
{4m_t^3}  + V(\bmr)
- (\sqrt{s}-2m_t-2\delta m_t(\nu)+i\Gamma_t)
\,\bigg)G(\bmr,\bmr^\prime,\sqrt{s},\nu) \, = \, 
\delta^{(3)}(\bmr-\bmr^\prime) 
\,,
\end{equation}
where $G$ is the Green function.\footnote{
The LL zero-distance Greens function in dimensional regularization 
has the simple analytic form\\
$
 G^0(0,0,\sqrt{s},\nu) =
 \frac{m_t^2}{4\pi}\left\{\,
 i\,v - a\left[\,\ln\left(\frac{-i\,v}{\nu}\right)
 -\frac{1}{2}+\ln 2+\gamma_E+\psi\left(1\!-\!\frac{i\,a}{2\,v}\right)\,\right]
 \,\right\}
 +\,\frac{m_t^2\,a}{4 \pi}\,\,\frac{1}{4\,\epsilon}
$, with
$a=C_F\alpha_s(m_t\nu)$ and
$v=((\sqrt{s}-2m_t-2\delta m_t(\nu)+i\Gamma_t)/m_t)^{1/2}$.
}
The Lagrangian does not describe $t\bar t$ production or annihilation. This is
done by additional operators (external currents). The dominant operators,
which describe $t\bar t$ production in a S-wave spin-triplet state are
\begin{eqnarray}
{\cal O}_{V,A\,\, \bmp}  =  C_{V,A}(\nu)
\left[\,\bar e\,\gamma_j(\gamma_5)\,e\,\right]\,
\left[\,\psi_{\bmp}^\dagger\, \sigma_j (i\sigma_2)
  \chi_{-\bmp}^*\,\right]
\,.
\end{eqnarray}
where the coefficients $C_{V,A}(\nu)$ describe the hard, nonresonant
fluctuation that are involved in the $t\bar t$ production process. 
Using the optical theorem one can obtain the total cross section,
\begin{eqnarray}
R \, = \, \frac{\sigma_{t\bar t}}{\sigma_{\mu^+\mu^-}} \, \sim \,
\mbox{Im}\left[\,(C_V^2(\nu)+C_A^2(\nu))\, G(0,0,\sqrt{s},\nu)
\,\right]
\,,
\label{optical}
\end{eqnarray}
where the zero-distance Green function describes the nonrelativistic
dynamics of the $e^+e^-\to e^+e^-$ forward scattering amplitude with
the $t\bar t$ pair being produced and annihilated at the origin. (In the
presentation here I neglect the effects from the subleading S- and P-wave
$t\bar t$
production operators that contribute at NNLL order. Their contributions are 
completely known~\cite{HMST2}.) The
zero-distance Green function is fully known at NNLL order (see
e.g. Ref.~\cite{HMST2} for details). The matching conditions
$C_{V,A}(1)$ 
are known at NNLL order. The renormalization evolution of the
coefficients, $f(\nu) = C_{V,A}(\nu)/C_{V,A}(1)$ is known at NLL
order~\cite{ultrasoft,Pineda1}. At NNLL order the 3-loop non-mixing
contributions are known for the function $f$ ~\cite{3loop}, however the subleading
evolution of the coefficients that go into the NLL evolution equation
of $f$ (mixing contributions) has not yet been determined apart from
relatively small spin-dependent contributions~\cite{Penin1}.
In Fig.~\ref{figttnnll} the predictions for 
$R$ at LL (blue dotted lines), NLL (green dashed lines) and NNLL order
without the mixing corrections (red solid lines) are shown for
$m_t=175$~GeV, $\Gamma_t=1.43$. We see that the
NNLL corrections are substantial, and that the NNLL prediction has a
much larger $\nu$-dependence than the NLL order one. At present the
QCD normalization uncertainty is
around~$6\%$~\cite{HoangEpiphany}. This is far 
away from the $3\%$ goal, but a final conclusion has to await the
completion of the missing NNLL order corrections or even higher order
computations. I believe that, in practice, it will likely be the full
NNLL order prediction that will determine the final QCD uncertainty that
can be achieved since a complete NNNLL order computation appears out
of reach. Partial results at this order are nevertheless already
available, see e.g. Refs.~\cite{Penin2,Beneke1,Penin3,Eiras1}. 

\section{Finite Lifetime and Electroweak Effects}
\label{sectionew}

Until now most effort in the literature went into the analysis and
determination of QCD effects. Electroweak and in particular finite top
quark lifetime corrections have receive much less attention beyond the
LL order level. In fact not even the full set of NLL order corrections,
based on the counting in Eq.~(\ref{powercounting}, are known for the
total cross section. For the treatment of finite lifetime effects
no fully general method, that can at the same time handle all realistic
cases and observables and can address all subtleties, exists. Within a
given set of (reasonable) approximations   
and for specific observables, however, a systematic and consistent approach
can be developed. In Ref.~\cite{HoangReisser1} the EFT approach
developed for the 
QCD effects was extended to determine the finite topquark lifetime
corrections to the total cross section for the
powercounting~(\ref{powercounting}) in the approximation of a stable $W$
boson. The approach is very similar to the 
theory of light propagation in an absorptive medium, where the effects 
of absorption can be encoded in complex contributions to the coefficients
of the vacuum theory as long as one does not want to address
microscopic details of the absorption processes. 

Electroweak effects can be  categorized into three classes:
\begin{itemize}
\item[(a)] "Hard" electroweak effects: This class includes hard,
  point-like electroweak effects related e.g. to the $t\bar t$
  production mechanism by virtual photon and Z exchange, or
  corrections to various matching conditions 
  of the EFT. In general these corrections are modifications of the
  hard QCD matching conditions of the EFT operators. They can be
  determined by standard methods through matching at the top quark
  complex pole, and are real numbers. 
\item[(b)] Electromagnetic effects: They are
  relevant for the luminosity spectrum of the $e^+e^-$ initial state
  and require the treatment of effects related to the collinear and
  ultrasoft fluctuations that come in when the photonic interactions
  with the $e^+e^-$ initial state are accounted for.
  The other low- and high energy properties of photon interactions of
  the $t\bar t$ pair and its decay products are similar to the gluonic
  corrections and can be incorporated in the same way into the
  nonrelativistic EFT, 
  but their effects are in general of higher order (see
  Eq.~(\ref{powercounting})). At NNLL order a coherent treatment 
  of all electromagnetic effects is required.
\item[(c)] Effects related to the finite top quark lifetime: Apart
  from the top decay (into $Wb$ for the Standard Model) this class
  also includes interference contributions with processes having the
  same $W^+W^-b\bar b$ final state but only one or even no top quark
  at intermediate stages. It also accounts for interactions involving
  the top decay products (sometimes called "nonfactorizable"
  effects). In Ref.~\cite{HoangReisser1} it was shown that, as long as the top
  decay is treated inclusively, finite lifetime effects can be
  incorporated into the EFT matching conditions by imaginary
  contribution that are determined from those (and only those) cuts in
  electroweak matching corrections that are related to the top
  decay. As for class (a) the matching procedure is carried out at the
  top quark complex pole (although this issue does not become relevant
  up to NNLL order). 
  One might say that the top decay is integrated out, although this notion can
  be misleading. This renders the EFT non-Hermitian, but unitarity is still 
  preserved due to the hermiticity of the full electroweak theory. 
  Gauge invariance is maintained at all times (as long as only gauge
  invariant sets of operators are used in the EFT).
\end{itemize}
Let me now discuss the status concerning the electroweak effects for
the total cross section coming from the three classes and using the
power counting~(\ref{powercounting}): 
At {\bf LL order} all contributions are known. There are the tree level
matching conditions $C_{V,A}(1)$ to the coefficients of the $t\bar t$
production and annihilation operators which describe the intermediate
virtual photon and Z exchange in $e^+e^-$ annihilation (class(a)), the
luminosity spectrum in Eq.~(\ref{convolution}) (class (b)) and the
imaginary width term $\psip{\bmp}^\dagger i\Gamma_t \psip{\bmp}$ in the
kinetic terms of the EFT Lagrangian in 
Eq.~(\ref{Lke}). The width term arises from the $Wb$ cuts in the full
theory top quark electroweak selfenergy diagrams in the matching
procedure.  
At {\bf NLL order} there are no corrections in class (a), because hard
electroweak matching corrections beyond LL order can only
contribute at NNLL order according to~(\ref{powercounting}). Moreover
the NNLL class (a,c) matching corrections to the Coulomb potential
${\cal V}_c$ and the dominant top-$A^0$ gluon interactions vanish due
to gauge invariance. In class (b) there is an additional QED
contribution to the Coulomb potential from the exchange of a Coulomb
photon (see~\cite{Signertalk}). A similar correction does not exist at NNLL
order. In class (c) there are ${\cal O}(\alpha_s)$ QCD corrections to
the top width $\Gamma_t$~\cite{Jezabek1} and phase space corrections which are
explained below. The statement I made on the class (a) corrections
implies a cancellation of effects that involve
real and virtual ultrasoft gluon exchange among the top quarks and its
decay products that can appear nontrivial from the diagrammatic point of
view, see e.g. Refs.~\cite{Melnikov1,Modritsch1}.

\begin{figure*}[t]
\begin{center}
{\includegraphics[bb=100 430 580 700,width=7.4cm]{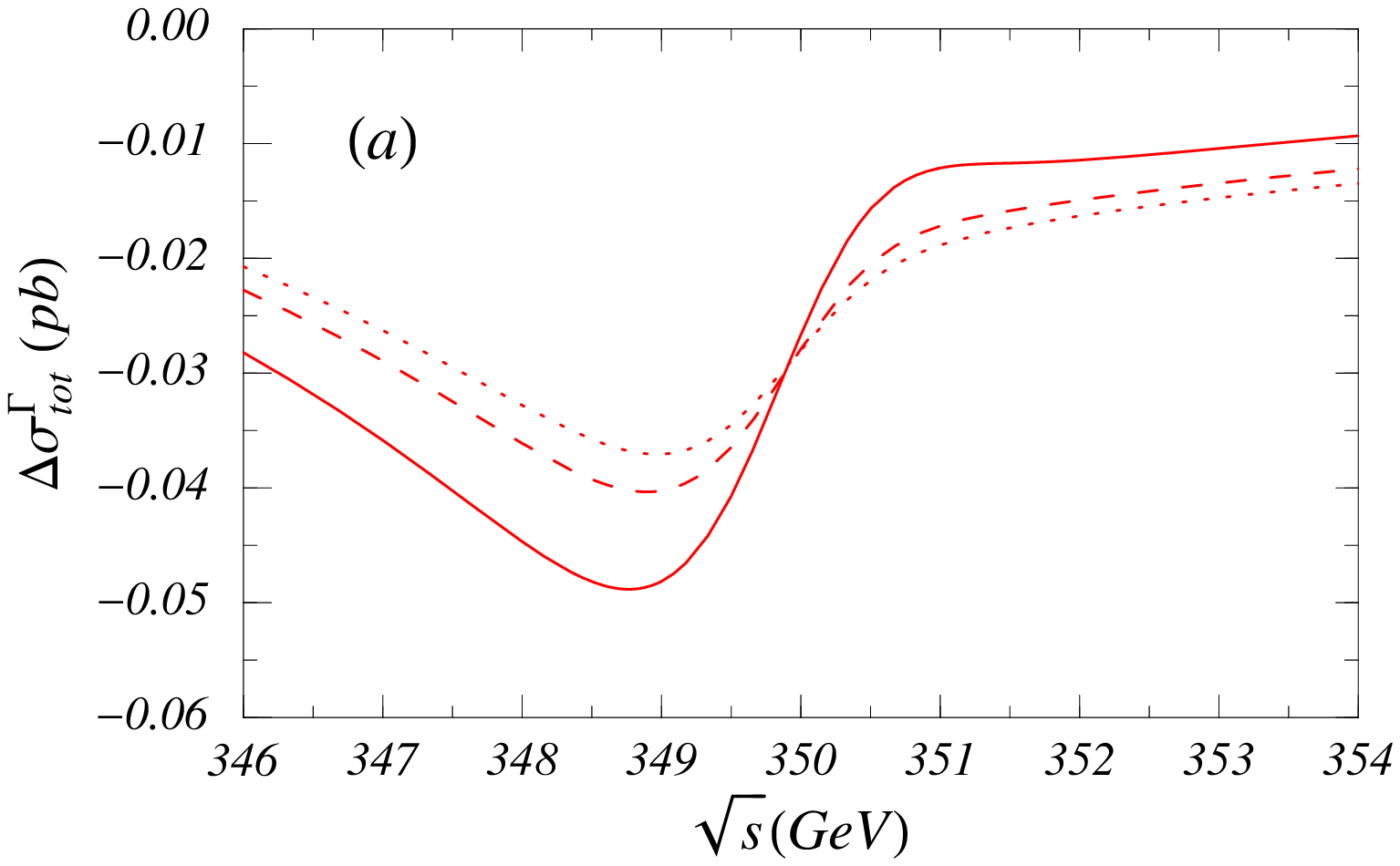}
\includegraphics[width=7.1cm]{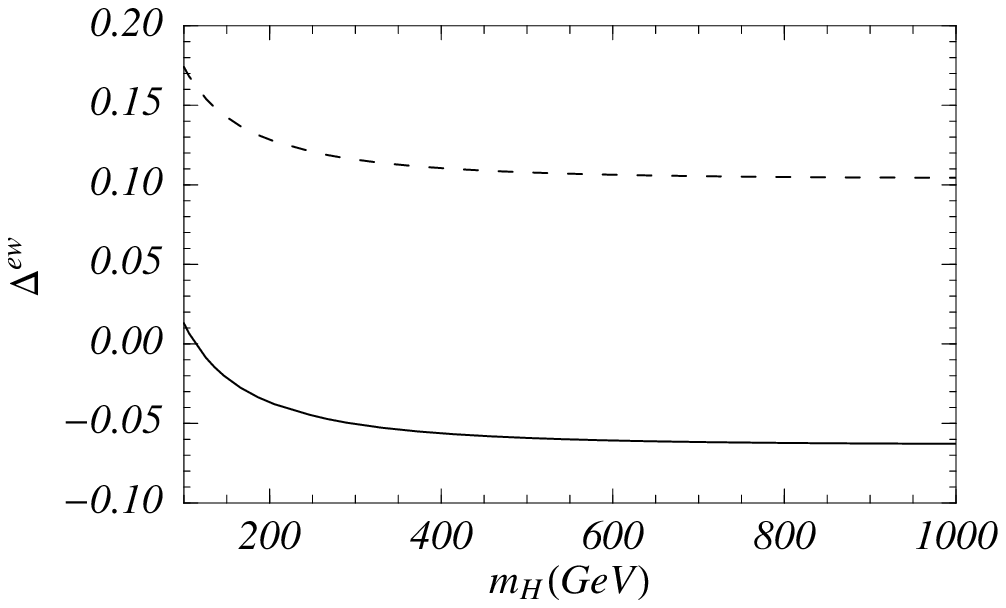}}
\caption{Left panel: Corrections to the total cross section from 
NNLL time-dilatation and interference effects and NLL summation of
phase space logarithms~\cite{HoangReisser1}.
 Right panel: Relative normalization corrections to the total cross
 section from NNLL hard electroweak corrections for $\alpha=1/137$
 (dashed line) and for a scheme with an electromagnetic coupling at
 the $m_t$ scale,
 $\alpha^{n_f=8}(\mu=m_t)=125.9$~\cite{HoangReisser2}. 
} 
\label{figew}
\end{center}
\end{figure*}
\begin{figure}[t] 
\includegraphics[width=6.7cm]{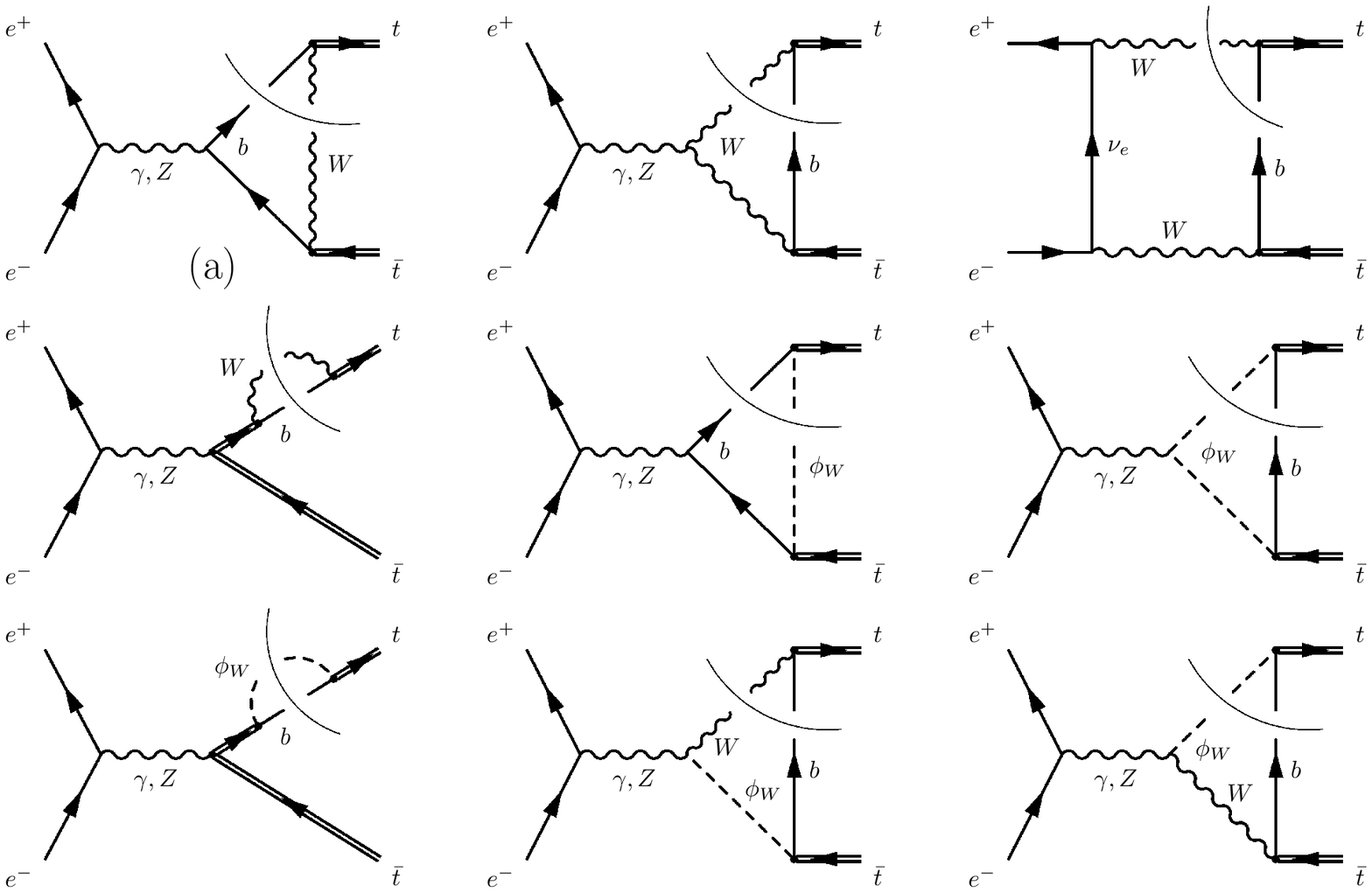}
\hspace{0.9cm}
\includegraphics[width=6.7cm]{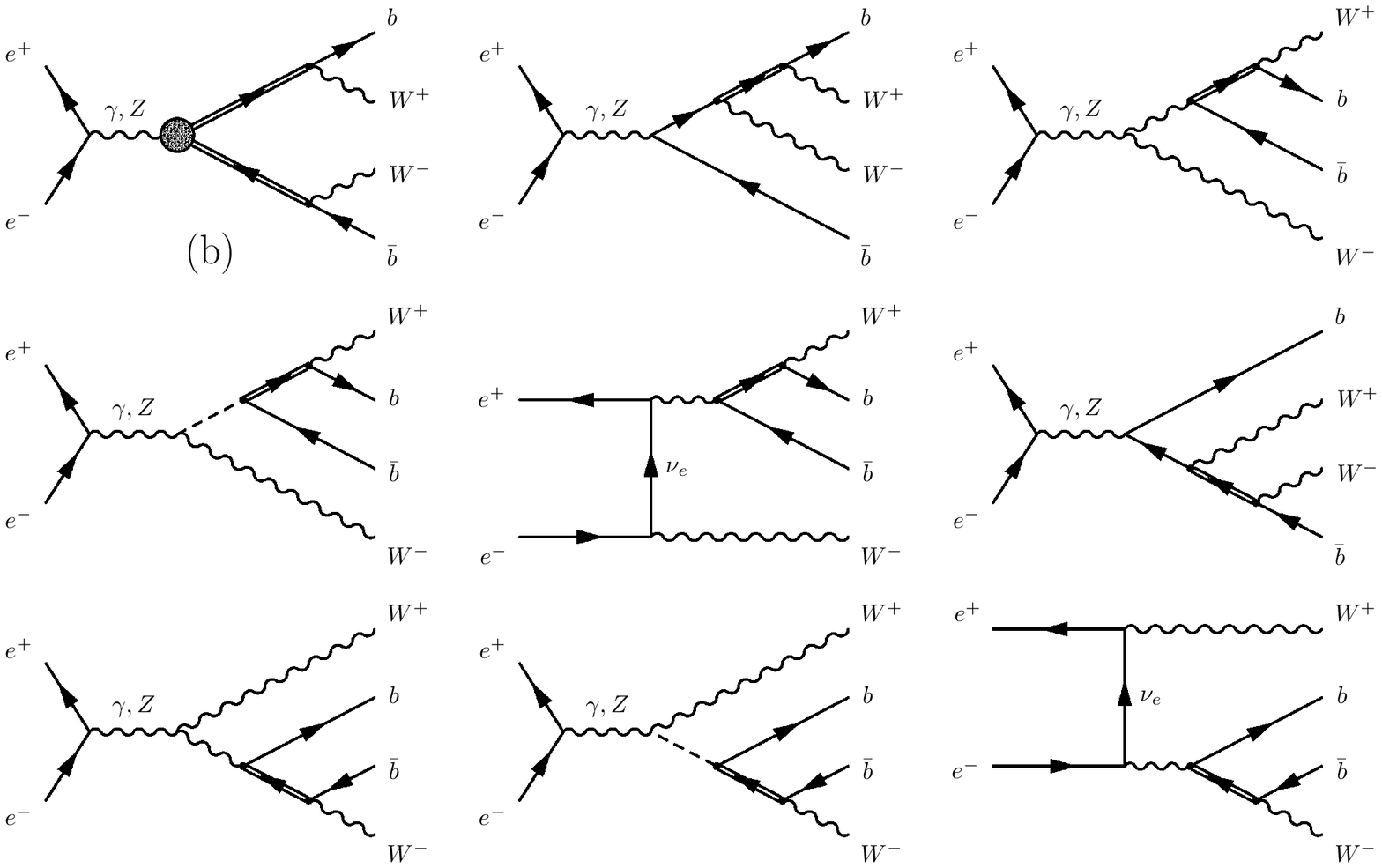}
 \caption{
Left panel: Full theory diagrams in Feynman gauge needed to determine the 
electroweak absorptive parts in the Wilson coefficients $C_{V/A}$  
related to the physical $bW^+$ and $\bar b W^-$ intermediate states. Only the
$bW^+$ cut is drawn explicitly. 
Right panel: Full theory diagrams describing the process 
$e^+e^-\to bW^+\bar b W^-$ with one or two intermediate top or antitop quark
propagators. The circle in the first diagram represents the QCD form
factors for the $t\bar t$ vector/axial-vector currents.
 \label{figewpics} }
\end{figure}

At {\bf NNLL order}, there are one-loop
electroweak corrections in class (a) to the matching conditions
$C_{V,A}(1)$. They were determined in
Refs~\cite{Grzadkowski,GuthKuehn} and updated
recently in Ref.~\cite{HoangReisser2}. They lead to an
energy-independent 
normalization correction to the total cross section that is shown in
Fig.~\ref{figew} as a function of the Higgs mass. The systematic and
consistent determination of NNLL order electromagnetic corrections has
not been attempted until now. Their determination represents a piece
of work from which one can learn a lot. 
The class (c) finite lifetime corrections were analyzed in
Ref.~\cite{HoangReisser1}. They include the ${\cal O}(\alpha_s^2)$ QCD
and one-loop electroweak corrections
to $\Gamma_t$~\cite{Blokland1,DennerSack}, the
time-dilatation correction 
$\propto\Gamma_t\bmp^2/m_t$ (see Eq.~(\ref{Lke})) and the one-loop
imaginary contributions to 
$C_{V,A}(1)$ from cuts related to the top decay in the one-loop
corrections to $e^+e^-\to t\bar t$ (see Fig.~\ref{figewpics}a).
Due to unitarity the imaginary contribution have the same sign for the
$t\bar t$ production and the annihilation operators and lead to a
energy-dependent correction to the total cross
section~(\ref{optical}). It was shown in Ref.~\cite{HoangReisser1} that these
corrections account for the interference of the dominant amplitude
$e^+e^-\to t\bar t \to b\bar b W^+W^-$ with amplitudes having the same
final state, but only one top or antitop as intermediate lines
(Fig.~\ref{figewpics}b). These interference corrections are an
integral part of the $t\bar t$ cross section once the top finite
lifetime is accounted for. They lead to a new
type of UV-divergence that can be seen from Eq.~(\ref{optical}) and the
UV-divergence in the real part of the zero-distance Green function,
seen footnote~1. The 
divergence is related to the Breit-Wigner-type EFT top quark
propagator, see Eq.~(\ref{Lke}), and the fact that the EFT phase space
is infinite. This is because the EFT is based on an expansion around
the top mass shell region, which in this context means taking 
$m_t\to\infty$. The UV-divergence has to be renormalized by imaginary
counterterms of $(e^+e^-)(e^+e^-)$ forward scattering
operators~\cite{HoangReisser1} which also have to be added to the RHS of
Eq.~(\ref{optical}). The renormalization group evolution of these
operators is a NLL order effect (just like the LL running of $\alpha_s$
is determined from NLL one-loop diagrams) and was determined in
Ref.~\cite{HoangReisser1}. It sums phase space logarithms $\propto \Gamma_t
(\alpha_s\ln v)^n$ to all orders in $\alpha_s$.
The matching conditions of the coefficients of the  $(e^+e^-)(e^+e^-)$
operators account for the difference between the infinite EFT phase
space and the physical one~\cite{inpreparation}.
The impact of the sum of time-dilatation and interference corrections,
and the NLL order phase space logarithms is shown in
Fig.~\ref{figew}a. The corrections are energy-dependent and
particularly large where the cross section is small and the $t\bar t$
invariant mass is below $2m_t$. They also leads to a shift of the
peak position in $\sigma^0_{t\bar t}$ by $30$ to $50$~MeV. The results in
Fig.~\ref{figew} demonstrate that the NLL and NNLL electroweak
corrections known by now are comparable to the corresponding QCD
effects, and that the full set of NLL and NNLL order electroweak
corrections needs to be determined to reach the $3\%$ goal
discussed at the beginning of this talk.

\section{Threshold Physics and $e^+e^-\to t\bar t H$}
\label{sectionyukawa}

\begin{figure}[htb] 
  \begin{center}
    \includegraphics[width=6cm]{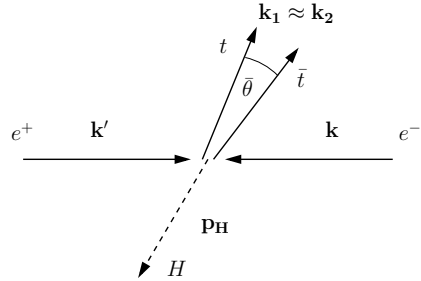}
    \vskip  0.0cm
    \caption{
      Typical constellation of momenta for the process $e^+e^-\to t\bar t H$ 
      in the large Higgs energy endpoint region. 
      \label{figHiggs} }
  \end{center}
\end{figure}

\begin{figure}[t!] 
\includegraphics[bb=80 450 560 730,width=7cm]{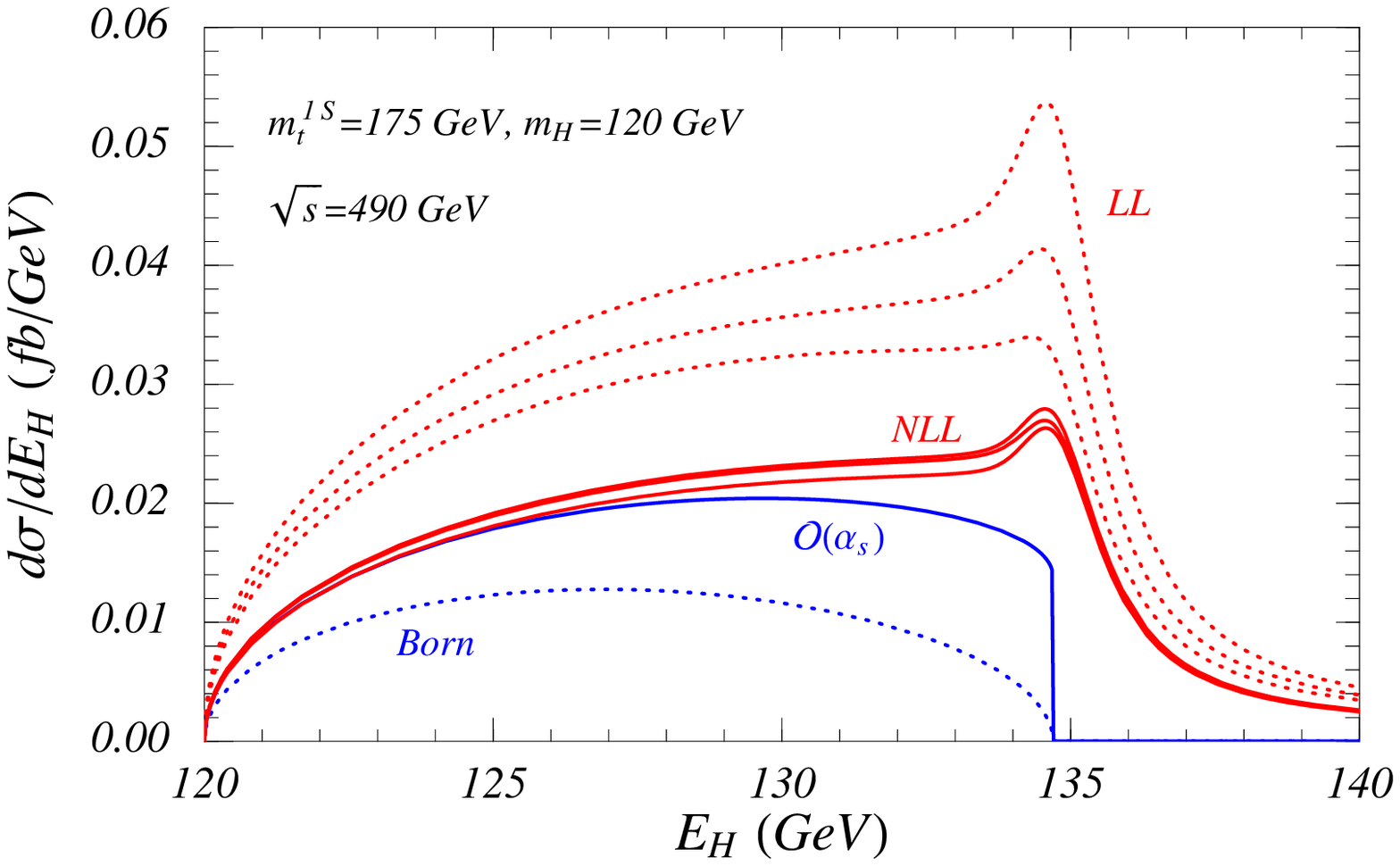}
\hspace{0.9cm}
\includegraphics[bb=80 450 560 730,width=7cm]{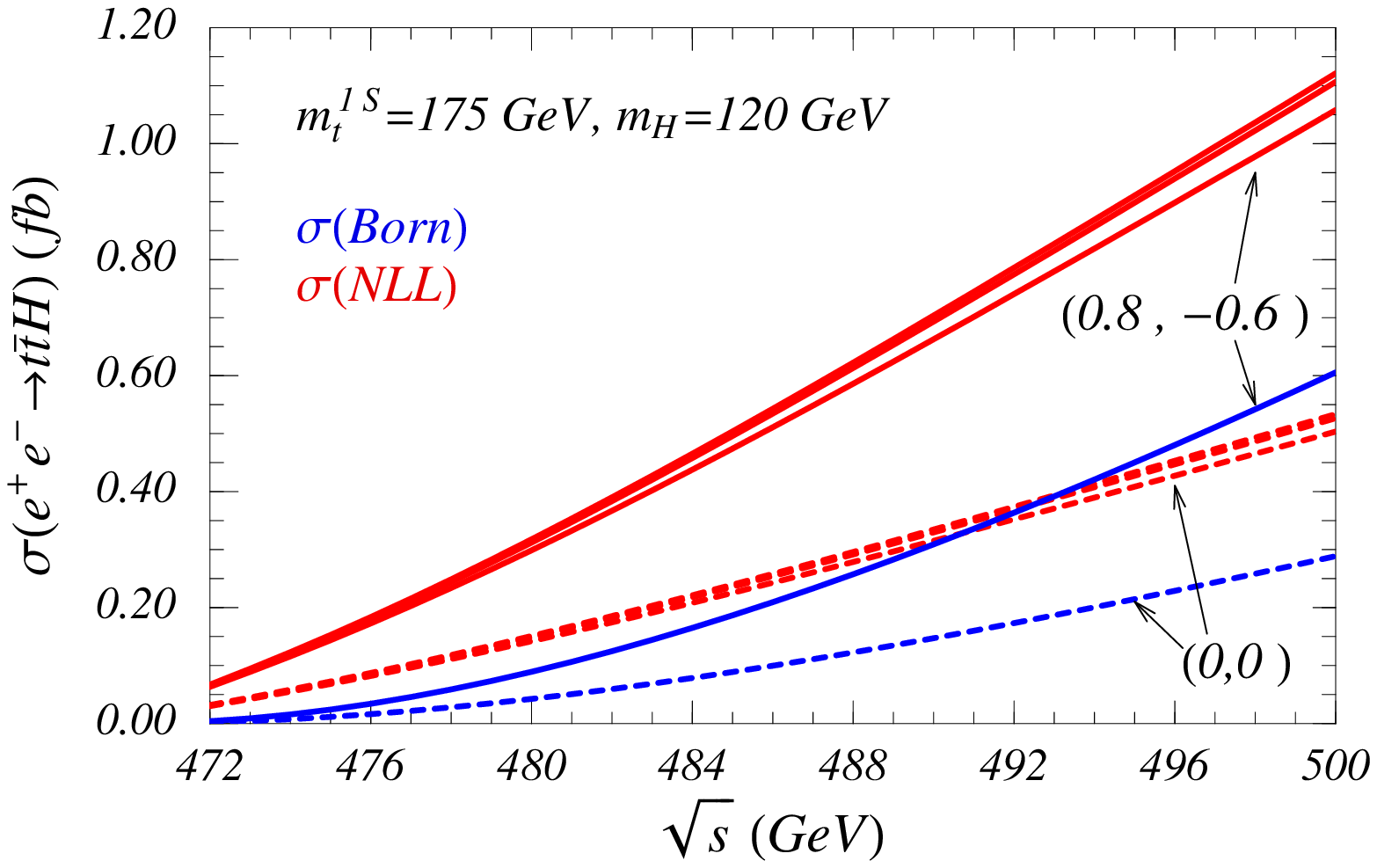}
 \caption{
Left panel: Higgs energy spectrum for the the process $e^+e^-\to t\bar t H$
at $\sqrt{s}=490$~GeV for $m_t=175$~GeV, $m_H=120$~GeV, $\Gamma_t=1.43$~GeV  
at NLL (red solid lines) and LL order (red dotted lines) for
$\nu=0.1,0.2,0.4$. Also displayed are the tree level and ${\cal O}(\alpha_s)$
results for stable top quarks. Right panel: Total cross section
$\sigma(e^+e^-\to t\bar t H)$ as a function of the c.m.\,energy for the same
parameter set at tree level (blue lines) and NLL order (red lines) for
unpolarized (dotted lines) and polarized $e^+e^-$ beams with 
$(P_{e^+},P_{e^-})=(0.8,-0.6)$ (solid lines).
 \label{figtth} }
\end{figure}

It is one of the major tasks of the future Linear Collider to unravel
details of the mechanism of electroweak symmetry breaking. One of the
crucial measurements is the (as much as possible) model-independent
determination of the top Yukawa coupling $\lambda_t$ which in the Standard
Model is related the top mass and the Higgs vacuum expectation value. At the
$e^+e^-$ Linear Collider the top quark Yukawa coupling 
can be measured from top quark pair production associated with a
Higgs boson, $e^+e^-\to t\bar t H$, since the process is dominated by
the amplitude describing Higgs radiation off the $t\bar t$ pair. This
process is particularly suited for a light Higgs boson because the cross
section can then reach the $1$-$2$~fb level and measurements of
$\lambda_t$ (close to the Standard Model value) with relative
errors of around $5\%$ are possible~\cite{TESLATDR,Gay1}. With this
motivation 
one-loop QCD~\cite{Dittmaier1,Dawson2} and also electroweak
corrections~\cite{Belanger1,Denner1,You1} were 
determined. There is, however, a region in the phase space where the
Higgs energy is large and the $t\bar t$ dynamics is nonrelativistic. For
large Higgs energies the $t\bar t$ pair is forced to become collinear
and to move opposite to the Higgs direction in order to achieve the
large total momentum necessary to balance the large Higgs momentum
(Fig.~\ref{figHiggs}).
%
%
 In this kinematic region the $t\bar t$
invariant mass $Q_{t\bar t}$ is close to $2m_t$, i.e.\,the $t\bar t$
pair is nonrelativistic in its c.m.~frame. For a relatively light
Higgs below the $W^+W^-$ threshold the Higgs width is only at the level of
several MeV and it is therefore possible to neglect gluon interactions between 
the top quarks and the Higgs decay products. So the QCD dynamics of
the $t\bar t H$ system in the large Higgs energy endpoint region is
very similar to the physics at the $t\bar t$ threshold discussed in
the previous sections. In particular, the usual loop QCD perturbation
theory breaks   down due to $(\alpha_s/v)^n$ and $(\alpha_s\ln v)^n$
singularities and an EFT treatment is required. In
Ref.~\cite{HoangFarrell1} a 
NLL order QCD factorization formula in close analogy to Eq.~(\ref{optical})
was derived for the Higgs energy spectrum in the large Higgs energy
endpoint region using the formalism developed for the $e^+e^-\to t\bar
t$ threshold. It has the (simplified) form
\begin{eqnarray}
\frac{d\sigma}{d E_H}(E_H\approx E_H^{\rm max}) & = &
h(\sqrt{s},m_t,m_H)\,
\,\mbox{Im}\left[\, C^2(\nu)\,G(0,0,Q_{t\bar t}(E_H),\nu)\,\right]\,
\,,
\label{dsdEHEFT}
\end{eqnarray}
where the constant $h$ accounts for the hard electroweak effects (where
arguments such the Z mass and the electroweak couplings are not written) and
$C(\nu)$ for the hard QCD corrections of the $t\bar t H$ production
mechanism. For c.m.\,energies above $500$~GeV the summation of the
terms singular in $v$ leads to corrections to the known 
${\cal O}(\alpha_s)$ one-loop predictions for the total cross section
since for large c.m.\,energies only a part of the phase space is
dominated by the nonrelativistic $t\bar t$ dynamics. Below $500$~GeV,
however, the energy available during the first phase of a LC program
based on the cold technology,
the maximal possible relative top velocity is so 
small that the full phase space is nonrelativistic, i.e.\,the physics
relevant at the large Higgs energy endpoint in fact governs the full
phase space. This 
makes the loop expansion in powers of $\alpha_s$ unreliable and the
nonrelativistic expansion based on Eq.~(\ref{powercounting}) has to be
applied. In Ref.~\cite{HoangFarrell2} the factorization
formula~(\ref{dsdEHEFT}) 
was extended to also account for the correct physical behavior at the
low Higgs energy endpoint $E_H=m_H$ and the resulting NLL order QCD
predictions were analyzed for $\sqrt{s}\le 500$~GeV. (The NLL order
electroweak and finite lifetime corrections are still unknown.) 
In Fig.~\ref{figtth}a the Higgs energy spectrum at LL (red dotted
lines) and NLL order (red solid lines) are shown for
$\sqrt{s}=490$~GeV and $m_t^{\rm 1S}=175$~GeV, $\Gamma_t=1.43$~GeV,
$m_H=120$~GeV and 
$\nu=0.1,0.2,0.4$. As a comparison also the tree level (blue dotted line)
and ${\cal O}(\alpha_s)$ (blue solid line) predictions (for
$\Gamma_t=0$) are shown. In Fig.~\ref{figtth}b the total cross section
is shown as a function of the c.m.\,energy for the same choice of the
other parameters as in  Fig.~\ref{figtth}a at NLL order (red lines)
and tree level (blue lines). The dashed lines are for unpolarized
$e^+e^-$ beams, $(P_{e^+},P_{e^-})=(0,0)$, and the solid lines for
$(P_{e^+},P_{e^-})=(0.8,-0.6)$. We see that the NLL predictions are
substantially larger than the tree level ones, by roughly a factor of
two for c.m.\,energies around $500$~GeV. The cross section can be even
further enhanced when polarized $e^+e^-$  
beams are used. Since the past experimental simulation analyses for
top Yukawa coupling measurements at c.m.\,energies up to
$500$~GeV were based on tree level theory predictions and
unpolarized beams\footnote{In the experimental simulation analysis of
Ref.~\cite{Justetalk} the prospect 
$(\delta \lambda_t/\lambda_t)^{\rm ex}= 25\%$ 
was  obtained.}, it can be expected that the NLL predictions will have a   
substantial (positive) impact on the prospects for top Yukawa coupling 
measurements at these energies. 

\section{Threshold Production of Squark Pairs}
\label{sectionsquark}

Many models of supersymmetry breaking predict that at least one of the
supersymmetric partners of the top quark is sufficiently light such
that stop-antistop pair production is possible at a future Linear
Collider running at c.m.\,energies up to 1~TeV. In such a scenario
threshold measurements in analogy to the $t\bar t$ threshold will be
possible~\cite{nowaktalk}. An important difference to the $t\bar t$ case is,
however, that squark pairs are predominantly produced in a P-wave in
$e^+e^-$ annihilation such that the rise of the total cross section at the
threshold is $v^2$-suppressed and substantially slower than for $t\bar t$
production. Up to now there have not even been
consistent LL order predictions for this P-wave process because, here,
the top quark finite lifetime issues
that became relevant at NLL and NNLL order for S-wave
production come in already at LL order. In Ref.~\cite{HoangFemania1} a
first step 
toward a systematic treatment of the squark pair threshold was done by
construction the scalar version of vNRQCD relevant for the description
of the NLL and NNLL QCD effects.  

\section*{Acknowledgements}
I would like to thank my collaborators C.~Farrell, C.~Rei\ss{}er and
P.~Ruiz-Femenia for their excellent work and enthusiasm. I thank A.~Manohar
I.~Stewart and T.~Teubner for their collaboration over many years.
I also thank
the organizers of International Workshop on Top Quark Physics for the
pleasant and inspiring atmosphere during the conference.


\begin{thebibliography}{99}


\bibitem{Boogert:2002jr}
  S.~T.~Boogert and D.~J.~Miller,
  arXiv:hep-ex/0211021.


\bibitem{Cinabro1}
D.~Cinabro,
arXiv:hep-ex/0005015.

\bibitem{boogerttalk}
 S.~T.~Boogert, talk given at the ECFA LC workshop, 
Durham, UK, Sept 1-4, 2004


\bibitem{TTbarsim}
M.~Martinez and R.~Miquel,
Eur.\ Phys.\ J.\ C {\bf 27}, 49 (2003)
[arXiv:hep-ph/0207315].

\bibitem{renormalons}
A.~H.~Hoang, M.~C.~Smith, T.~Stelzer and S.~Willenbrock,
Phys.\ Rev.\  {\bf D59}, 114014 (1999)
[hep-ph/9804227];
%
M.~Beneke,
Phys.\ Lett.\  {\bf B434}, 115 (1998)
[hep-ph/9804241].


\bibitem{synopsis}
A.~H.~Hoang {\it et al.},
in Eur.\ Phys.\ J.\ direct C {\bf 2}, 1 (2000)
[arXiv:hep-ph/0001286].


\bibitem{1Smass}
A.~H.~Hoang, Z.~Ligeti and A.~V.~Manohar,
Phys.\ Rev.\ Lett.\  {\bf 82}, 277 (1999)
[hep-ph/9809423];
%
Phys.\ Rev.\  {\bf D59}, 074017 (1999)
[hep-ph/9811239].
%

\bibitem{HT1S}
 A.~H.~Hoang and T.~Teubner,
  Phys.\ Rev.\ D {\bf 60}, 114027 (1999)
  [arXiv:hep-ph/9904468].

%
\bibitem{Peralta1}
D. Peralta, M. Martinez and R. Miquel, talk presented at the
{\it 4th International Workshop on Linear Colliders}, 
Sitges, Barcelona, Spain, April 28 - May 5 1999.
%

\bibitem{BBL}
  G.~T.~Bodwin, E.~Braaten and G.~P.~Lepage,
  Phys.\ Rev.\ D {\bf 51}, 1125 (1995)
  [Erratum-ibid.\ D {\bf 55}, 5853 (1997)]
  [arXiv:hep-ph/9407339].

\bibitem{pNRQCD}
  N.~Brambilla, A.~Pineda, J.~Soto and A.~Vairo,
  Nucl.\ Phys.\ B {\bf 566}, 275 (2000)
  [arXiv:hep-ph/9907240].

\bibitem{LMR}
  M.~E.~Luke, A.~V.~Manohar and I.~Z.~Rothstein,
  Phys.\ Rev.\ D {\bf 61}, 074025 (2000)
  [arXiv:hep-ph/9910209].

\bibitem{NRQCDc}
  S.~Fleming, I.~Z.~Rothstein and A.~K.~Leibovich,
  Phys.\ Rev.\ D {\bf 64}, 036002 (2001)
  [arXiv:hep-ph/0012062].

\bibitem{ultrasoft}
  A.~H.~Hoang and I.~W.~Stewart,
  Phys.\ Rev.\ D {\bf 67}, 114020 (2003)
  [arXiv:hep-ph/0209340].


\bibitem{ManoharSoto1}
  A.~V.~Manohar, J.~Soto and I.~W.~Stewart,
  Phys.\ Lett.\ B {\bf 486}, 400 (2000)
  [arXiv:hep-ph/0006096].



\bibitem{Signertalk}
A.~Signer, talk given at this workshop.


\bibitem{HMST2}
  A.~H.~Hoang, A.~V.~Manohar, I.~W.~Stewart and T.~Teubner,
  Phys.\ Rev.\ D {\bf 65}, 014014 (2002)
  [arXiv:hep-ph/0107144].

\bibitem{Pineda1}
  A.~Pineda,
  Phys.\ Rev.\ D {\bf 66}, 054022 (2002)
  [arXiv:hep-ph/0110216].

\bibitem{3loop}
  A.~H.~Hoang,
  Phys.\ Rev.\ D {\bf 69}, 034009 (2004)
  [arXiv:hep-ph/0307376].


\bibitem{Penin1}
  A.~A.~Penin, A.~Pineda, V.~A.~Smirnov and M.~Steinhauser,
  Nucl.\ Phys.\ B {\bf 699}, 183 (2004)
  [arXiv:hep-ph/0406175].


\bibitem{HoangEpiphany}
  A.~H.~Hoang,
  Acta Phys.\ Polon.\ B {\bf 34}, 4491 (2003)
  [arXiv:hep-ph/0310301].


\bibitem{Penin2}
  A.~A.~Penin and M.~Steinhauser,
  Phys.\ Lett.\ B {\bf 538}, 335 (2002)
  [arXiv:hep-ph/0204290].

\bibitem{Beneke1}
  M.~Beneke, Y.~Kiyo and K.~Schuller,
  Nucl.\ Phys.\ B {\bf 714}, 67 (2005)
  [arXiv:hep-ph/0501289].

\bibitem{Penin3}
  A.~A.~Penin, V.~A.~Smirnov and M.~Steinhauser,
  Nucl.\ Phys.\ B {\bf 716}, 303 (2005)
  [arXiv:hep-ph/0501042].

\bibitem{Eiras1}
  D.~Eiras and M.~Steinhauser,
  JHEP {\bf 0602}, 010 (2006)
  [arXiv:hep-ph/0512099].



\bibitem{HoangReisser1}
  A.~H.~Hoang and C.~J.~Reisser,
  Phys.\ Rev.\ D {\bf 71}, 074022 (2005)
  [arXiv:hep-ph/0412258].


\bibitem{Jezabek1}
M.~Jezabek and J.~H.~K\"uhn,
Nucl.\ Phys.\ B {\bf 314}, 1 (1989).


\bibitem{Melnikov1}
K.~Melnikov and O.~I.~Yakovlev,
Phys.\ Lett.\ B {\bf 324}, 217 (1994)
[arXiv:hep-ph/9302311].

\bibitem{Modritsch1}
  W.~Modritsch and W.~Kummer,
  Nucl.\ Phys.\ B {\bf 430}, 3 (1994).


\bibitem{Grzadkowski}
  B.~Grzadkowski, J.~H.~Kuhn, P.~Krawczyk and R.~G.~Stuart,
  Nucl.\ Phys.\ B {\bf 281}, 18 (1987).

\bibitem{GuthKuehn}
R.~J.~Guth and J.~H.~K\"uhn,
Nucl.\ Phys.\ B {\bf 368}, 38 (1992).



\bibitem{HoangReisser2}
  A.~H.~Hoang and C.~J.~Reisser,
  arXiv:hep-ph/0604104.


\bibitem{Blokland1}
I.~Blokland, A.~Czarnecki, M.~Slusarczyk and F.~Tkachov,
Phys.\ Rev.\ Lett.\  {\bf 93}, 062001 (2004)
[arXiv:hep-ph/0403221].

\bibitem{DennerSack}
  A.~Denner and T.~Sack,
  Nucl.\ Phys.\ B {\bf 358}, 46 (1991).


\bibitem{inpreparation}
 A.~H.~Hoang, C.~J.~Reisser and  P.~Ruiz-Femenia, work in preparation.


\bibitem{TESLATDR}
J.~A.~Aguilar-Saavedra {\it et al.}  [ECFA/DESY LC Physics Working Group
                  Collaboration],
arXiv:hep-ph/0106315.

\bibitem{Gay1}
  A.~Gay,
  arXiv:hep-ph/0604034.


\bibitem{Dittmaier1}
S.~Dittmaier, M.~Kramer, Y.~Liao, M.~Spira and P.~M.~Zerwas,
Phys.\ Lett.\ B {\bf 441}, 383 (1998)
[arXiv:hep-ph/9808433].




\bibitem{Dawson2}
S.~Dawson and L.~Reina,
Phys.\ Rev.\ D {\bf 59}, 054012 (1999)
[arXiv:hep-ph/9808443].


\bibitem{Belanger1}
G.~Belanger {\it et al.},
Phys.\ Lett.\ B {\bf 571}, 163 (2003)
[arXiv:hep-ph/0307029].


\bibitem{Denner1}
A.~Denner, S.~Dittmaier, M.~Roth and M.~M.~Weber,
Nucl.\ Phys.\ B {\bf 680}, 85 (2004)
[arXiv:hep-ph/0309274].


\bibitem{You1}
Y.~You, W.~G.~Ma, H.~Chen, R.~Y.~Zhang, S.~Yan-Bin and H.~S.~Hou,
Phys.\ Lett.\ B {\bf 571}, 85 (2003)
[arXiv:hep-ph/0306036].

\bibitem{HoangFarrell1}
  C.~Farrell and A.~H.~Hoang,
  Phys.\ Rev.\ D {\bf 72}, 014007 (2005)
  [arXiv:hep-ph/0504220].

\bibitem{HoangFarrell2}
  C.~Farrell and A.~H.~Hoang,
  arXiv:hep-ph/0604166.


\bibitem{Justetalk}
A.~Juste talk presented at the Chicago Linear Collider Workshop, Chicago, USA,
January 2002, 


\bibitem{nowaktalk}
 H.~Nowak, talks given at the ECFA LC workshop, 
Durham, UK, Sept 1-4, 2004 and ILWS 05, Stanford, USA,
18-22 March, 2005. 

\bibitem{HoangFemania1}
  A.~H.~Hoang and P.~Ruiz-Femenia,
  Phys.\ Rev.\ D {\bf 73}, 014015 (2006)
  [arXiv:hep-ph/0511102].


\end{thebibliography}
\end{document}